\documentstyle{mn2e}
\input psfig

  \makeatletter
  \ifx\CUP@mtlplain@loaded\undefined  
  \makeatother  
    
  \else
    
  \fi

\loadboldmathitalic 
\loadboldgreek      
  
\makeatletter
\ifx\CUP@mtlplain@loaded\undefined  
  \newfont\bit{cmbxti10 at 9pt}
\else
  \newfont\bit{mtbxti10 at 9pt}
\fi
\makeatother

\def\LaTeX{L\kern-.36em\raise.3ex\hbox{a}\kern-.15em
    T\kern-.1667em\lower.7ex\hbox{E}\kern-.125emX}

\newcommand{\gsim}{\mathrel{\hbox{\rlap{\lower.55ex \hbox {$\sim$}}
                   \kern-.3em \raise.4ex \hbox{$>$}}}}
\newcommand{\lsim}{\mathrel{\hbox{\rlap{\lower.55ex \hbox {$\sim$}}
                   \kern-.3em \raise.4ex \hbox{$<$}}}}

\title[The formation of brown dwarfs]{The formation mechanism of brown dwarfs}

\author[M. R. Bate et~al.]
  {Matthew R. Bate,$^{1,2}$\thanks{E-mail: mbate@astro.ex.ac.uk}
  Ian A. Bonnell,$^3$
  and Volker Bromm.$^{2,4}$\\
  $^1$School of Physics, University of Exeter, Stocker Road,
    Exeter EX4 4QL \\
  $^2$Institute of Astronomy, University of Cambridge, Madingley Road,
    Cambridge CB3 0HA \\
  $^3$School of Physics and Astronomy, University of St Andrews, North Haugh, St Andrews, Fife, KY16 9SS \\
  $^4$Harvard-Smithsonian Center for Astrophysics, 60 Garden Street, Cambridge,
MA 02138, U.S.A.
}

\date{Accepted for publication in MNRAS}

\begin{document}

\maketitle

\begin{abstract}
  We present results from the first hydrodynamical star formation
  calculation to demonstrate that brown dwarfs are a natural and 
  frequent product of the collapse and fragmentation of a 
  turbulent molecular cloud.  The brown dwarfs form via the
  fragmentation of dense molecular gas in unstable multiple 
  systems and are ejected from the dense gas before they have 
  been able to accrete to stellar masses.  
  Thus, they can be viewed as `failed stars'.  Approximately three
  quarters of the brown dwarfs form in gravitationally-unstable
  circumstellar discs while the remainder form in collapsing 
  filaments of molecular gas.  These formation mechanisms are very efficient, 
  producing roughly the same number of brown dwarfs as stars,
  in agreement with recent observations.
  However, because close dynamical interactions are involved in 
  their formation, we find a very low frequency of binary brown 
  dwarf systems ($\lsim 5$\%) and that those binary brown dwarf 
  systems that do exist must be close $\lsim 10$ AU.  Similarly, we find that
  young brown dwarfs with large circumstellar discs (radii $\gsim 10$ AU)
  are rare ($\approx 5$\%).
\end{abstract}


\section{Introduction}

The existence of brown dwarfs was incontrovertibly demonstrated
for the first time by the discovery of Gliese 229B \cite{Nakajimaetal1995}, 
a cool brown dwarf orbiting an M-dwarf.  
In the same year, other candidates later confirmed to be 
free-floating brown dwarfs were 
announced (e.g.\ Teide 1 by Rebolo, Zapatero-Osorio \& Martin 1995), 
along with PPl 15
which was later discovered to be a binary brown dwarf
(Basri \& Martin 1999).  
Observations now suggest
that brown dwarfs are as common as stars, although stars
dominate in terms of mass (e.g.\ Reid et al.\ 1999).  

Despite the abundance of brown dwarfs, their formation mechanism is currently
a mystery.  The typical thermal Jeans mass in molecular cloud cores
is $\approx 1$ M$_\odot$ (Larson 1999 and references therein).
Thus, the gravitational collapse of these cores might be expected 
to form stars, not brown dwarfs.

There are two obvious routes by which brown dwarf systems
(i.e.\ brown dwarfs without stellar companions) may form.
First, they may result from the collapse of low-mass cores 
(masses $\lsim 0.1$ M$_\odot$) that
are smaller (radii $\lsim 0.05$ pc) and denser 
($n$(H$_2$)$\gsim 10^7$ cm$^{-3}$)
than the cores that are typically observed (i.e.\ they have low masses
yet are still Jeans unstable).  
Thus, brown dwarfs would be `low-mass stars'.  Such low-mass bound 
cores have not yet been observed, but they would be difficult 
to detect because of their small sizes and low masses, and rarely 
detected due to their short dynamical time-scales ($\sim 10^4$ years).
Observations are beginning to reach this mass regime 
(e.g.\ Motte, Andr\'e \& Neri 1998), although the low-mass clumps found 
thus far probably are not gravitationally bound \cite{Johnstoneetal2000}.

The second possibility is that brown dwarfs form in higher-mass
cores but are prevented from accreting enough mass
to exceed the hydrogen-burning limit.  If such 
a core fragments to form an unstable multiple system, this 
may be achieved by the dynamical ejection of a fragment
from the core, cutting it off from the reservoir of gas,
and thus preventing it from accreting to a stellar mass.
In this case, brown dwarfs would be `failed stars'.
This ejection mechanism has been proposed by 
Reipurth \& Clarke \shortcite{ReiCla2001} and
Watkins et al.\ \shortcite{Watkinsetal1998b}.

In this letter, we present results from the first hydrodynamical 
calculation to demonstrate that a large number of brown dwarfs
can be formed during the fragmentation of a molecular 
cloud. All the brown dwarfs are formed by the ejection of fragments
from unstable multiple systems.
In section 2, we briefly
describe the numerical method and the initial conditions for
our calculation.  In section 3, we present results from our
calculation and compare them with observations.  Finally, in 
section 4, we give our conclusions.

\section{Computational method and initial conditions}

The calculation presented here was performed using a three-dimensional, 
smoothed particle hydrodynamics (SPH) code
based on a version originally developed by Benz (Benz 1990; 
Benz et al.~1990).  The smoothing lengths of particles vary in 
time and space, such that the number of 
neighbours for each particle remains approximately constant 
at $N_{\rm neigh}=50$.  We use the standard form of artificial viscosity 
(Monaghan \& Gingold 1983) with 
strength parameters $\alpha_{\rm_v}=1$ and $\beta_{\rm v}=2$.
Further details can be found in Bate, Bonnell \& Price 
\shortcite{BatBonPri1995}.  The code has been parallelised by M. Bate 
using OpenMP.

\subsection{Opacity limit for fragmentation and the equation of state}

When the collapse of a molecular cloud core begins,
the gravitational potential energy released easily 
radiates away so that the collapsing gas is approximately 
isothermal (e.g.\ Larson 1969).  Thus, the thermal pressure 
varies with density $\rho$ as $p \propto \rho^\eta$ where the 
effective polytropic exponent, 
$\eta \equiv {\rm dlog}[p]/{\rm dlog}[\rho] \approx 1$.
Fragmentation is allowed because the Jeans 
mass decreases with increasing density if $\eta <4/3$.

The opacity limit for fragmentation occurs when the
rate at which energy is released by the collapse exceeds the
rate at which the gas can cool (Low \& Lynden-Bell 1976; Rees 1976).  
The gas then heats up with $\eta > 4/3$, the Jeans mass increases,
and a Jeans-unstable collapsing clump quickly becomes Jeans-stable 
so that a pressure-supported fragment forms.  The gas begins to heat 
significantly at a density of $\approx 10^{-13}~{\rm g~cm}^{-3}$ 
\cite{MasInu2000}.

The pressure-supported fragment initially contains several Jupiter-masses 
(M$_{\rm J}$) and has a radius of 
$\approx 5$ AU \cite{Larson1969}.  Such a fragment is expected to be
embedded within a collapsing envelope; thus, its mass grows
with time.  Although these fragments later undergo another phase of 
collapse due to the dissociation of molecular hydrogen 
\cite{Larson1969} in order to become stars (or brown dwarfs), 
they are unlikely to sub-fragment (e.g.\ Boss 1989; Bate 1998, 2002).
Thus, the opacity limit sets a minimum fragment mass of 
$\approx 10$ M$_{\rm J}$ \cite{Boss1988}.  

To model the opacity limit for fragmentation without performing 
full radiative transfer, we use a gas equation of state given by 
$p = K \rho^{\eta}$,
where $K$ is a measure of the entropy of the gas.  The value of $\eta$ 
varies with density as
\begin{equation}\label{eta}
\eta = \cases{\begin{array}{rl}
1, & \rho \leq 10^{-13}~ {\rm g~cm}^{-3}, \cr
7/5, & \rho > 10^{-13}~ {\rm g~cm}^{-3}. \cr
\end{array}}
\end{equation}
We take the mean molecular weight of the gas to be $\mu = 2.46$.
The value of 
$K$ is defined such that when the gas is 
isothermal $K=c_{\rm s}^2$, with 
$c_{\rm s} = 1.84 \times 10^4$ cm s$^{-1}$ at 10 K,
and the pressure is continuous when the value of $\eta$ changes.

This equation of state reproduces the temperature-density relation
of molecular gas during spherically-symmetric collapse 
(as calculated with frequency-dependent radiative transfer)
to an accuracy of better than 20\% in the non-isothermal 
regime up to densities of 
$10^{-8}~{\rm g~cm}^{-3}$ \cite{MasInu2000}.  
Thus, our equation of state should model 
collapsing regions well, but may not model the equation of 
state in protostellar discs particularly accurately due to
their departure from spherical symmetry.

\subsection{Sink particles}

The opacity limit results in the formation of distinct 
pressure-supported fragments.  
As these fragments accrete, their central density 
increases, and it becomes computationally impractical 
to follow their internal evolution due to the short dynamical 
time-scales involved.  Therefore, when the central density 
of a fragment exceeds $10^{-11}~{\rm g~cm}^{-3}$, we insert 
a sink particle into the calculation (Bate et al.\ 1995).  
Gas within radius $r_{\rm acc}=5$ AU of the fragment's centre
(i.e.\ the location of the SPH particle with the highest density)
is replaced by a point mass with the same mass and momentum.
The fragment is only replaced if it is gravitationally bound.
Any gas that later falls within this radius is accreted by the 
point mass if it is bound.  Thus, we can only resolve discs 
around sink particles if they have radii $\gsim 10$ AU.  

Since all sink particles are created from the pressure-supported 
fragments, their initial masses are $\approx 10$ 
M$_{\rm J}$, as given by the opacity limit for fragmentation.
Subsequently, they may accrete large amounts of material 
to become higher-mass brown dwarfs ($\lsim 75$ M$_{\rm J}$) or 
stars ($\gsim 75$ M$_{\rm J}$).

The gravitational acceleration between two sink particles is Newtonian 
for $r \geq 4$ AU, but is softened within this radius using spline 
softening \cite{Benz1990}.  The maximum acceleration occurs at a 
distance of $\approx 1$ AU; therefore, this is the minimum binary 
separation that can be resolved.

\begin{figure*}
\caption{\label{pic1} Top panels: Time sequence showing the formation of two brown dwarfs via disc fragmentation.  Spiral density waves in a gravitationally-unstable circumbinary disc (panel 1; $t=5160$ years) fragment to form a third star (panel 2; $t=5735$ years) and two brown dwarfs (panel 3; $t=6020$ years).  The resulting multiple system undergoes chaotic evolution.  The first brown dwarf (box; red lines) is eventually ejected from the system, while the second is put into a highly eccentric orbit (panel 4; $t=10490$ years).  Much later the second brown dwarf (triangle; blue lines) is also ejected (panel 5; $t=17340$ years) leaving a hierarchical triple system.  Each panel is 600 AU across.  The colour scale shows the logarithm of column density ranging from $3 \times 10^{-1}$ to $3\times 10^3$ g~cm$^{-2}$.  Lower panels: The trajectories and masses of the five objects during the ejection of the two brown dwarfs.  The left panel has the same orientation as the time sequence.  The centre panel shows the third dimension.  The right panel shows the evolution of the masses of the objects.  When the brown dwarfs are ejected from the dense gas accretion on to them is halted and they are consequently unable to attain stellar masses.  Times are given in years from the beginning of star formation in the cloud. }
\end{figure*}

\begin{figure*}
\caption{\label{pic2} Time sequence showing the formation of two brown dwarfs via the fragmentation of collapsing filaments of molecular gas.  Two brown dwarfs (box, triangle; red and blue lines) fragment out of filaments (panels 1, 2, \& 3; $t=14300$, $t=14870$ and $t=15440$ years).  Both fall towards an existing multiple system, one almost radially along its filament (triangle).  The first brown dwarf (square; red line) is quickly ejected from the multiple system (panels 3 \& 4; $t=15440$ and $t=17150$ years).  Much later, after several crossing times of the multiple system, the second brown dwarf (triangle; blue line) is also ejected (panel 5; $t=23710$ years).  Each panel is 1000 AU across.  The colour scale shows the logarithm of column density ranging from $3 \times 10^{-1}$ to $3\times 10^3$ g~cm$^{-2}$.  Lower panels: The trajectories and masses of the objects during the ejection of the two brown dwarfs.  The left panel has the same orientation as the time sequence.  The centre panel shows the third dimension.  The right panel shows the evolution of the masses of the objects.  When the brown dwarfs are ejected from the dense gas, accretion onto them is halted and they are consequently unable to attain stellar masses.  Times are given in years from the beginning of star formation in the cloud.  For clarity, only the trajectories of the two brown dwarfs are coloured.}
\end{figure*}

\subsection{Initial conditions}

The initial conditions consist of a large-scale, turbulent 
molecular cloud.  The cloud is spherical and
uniform in density with a mass of 50 M$_\odot$ and
a diameter of $0.375$ pc.  At the temperature of 10 K,
the mean thermal Jeans mass is 1 M$_\odot$ 
(i.e.\ the cloud contains 50 thermal Jeans masses).
An initial supersonic turbulent velocity field is imposed on the
cloud in the same manner 
as Ostriker, Stone \& Gammie \shortcite{OstStoGam2001}.  We generate a
divergence-free random Gaussian velocity field with a power spectrum 
$P(k) \propto k^{-4}$, where $k$ is the wavenumber.  
In three dimensions, this gives a 
velocity dispersion that varies with distance, $\lambda$, 
as $\sigma(\lambda) \propto \lambda^{1/2}$ in agreement with the 
observed Larson scaling relations for molecular clouds \cite{Larson1981}.
The velocity field is normalised so that the kinetic energy of the 
turbulence equals the magnitude of the cloud's gravitational 
potential energy.

\subsection{Resolution}

The local Jeans mass must be resolved throughout 
the calculation to correctly model fragmentation (Bate \& Burkert 1997;
Truelove et al.\ 1997; Whitworth 1998).  Bate \& Burkert 
\shortcite{BatBur1997} found that this requires 
$\gsim 2 N_{\rm neigh}$ SPH particles per Jeans mass; 
$N_{\rm neigh}$ is insufficient.
We have repeated their calculation using different numbers of
particles and find that $1.5 N_{\rm neigh}=75$
particles is also sufficient
to resolve fragmentation (Bate, Bonnell \& Bromm, in preparation).
The minimum Jeans mass in the calculation presented here occurs 
at the maximum density during the isothermal phase of the 
collapse, $\rho = 10^{-13}$ g~cm$^{-3}$, 
and is $\approx 0.0011$ M$_\odot$ (1.1 M$_{\rm J}$).  Thus, we
use $3.5 \times 10^6$ particles to model the 50-M$_\odot$ cloud.
The calculation required approximately 95000 CPU hours on the SGI Origin 3800 
of the United Kingdom Astrophysical Fluids Facility (UKAFF).

\section{Results}
\label{evolution}

\subsection{Evolution of the cloud}

The hydrodynamical evolution of the cloud produces shocks which 
decrease the turbulent energy that initially supported
the cloud.  In parts of the cloud, gravity begins to dominate 
and dense self-gravitating cores form and collapse.  These dense cores are 
the sites where the formation of stars and brown dwarfs occurs.  
In all, 23 stars and 18 brown dwarfs are produced.
An additional 9 objects have sub\-stellar masses when we stop the
calculation, but are still accreting.  Three of these 
are likely to remain substellar, but the other six would
probably become stars.  The cloud's evolution and the properties of 
the stars and brown dwarfs will be discussed in detail in subsequent 
papers.  In this letter, we concentrate on the mechanism 
by which the brown dwarfs form.

\subsection{The formation of brown dwarfs}

Although the thermal Jeans mass 
is initially 1 M$_\odot$, the cloud produces 
roughly equal numbers of brown dwarfs
and stars.  By what mechanism(s) do these brown dwarfs form?

As mentioned in Section 2.2, 
all the stars and brown dwarfs begin as fragments whose initial 
masses are given by the opacity limit for fragmentation.
Thus, {\it all the objects begin with masses of a few Jupiter masses}.
Those that become stars subsequently accrete large quantities of gas 
from the dense cores in which they form, while those that remain as 
brown dwarfs do not.

Without exception, we find that those objects that end up as 
brown dwarfs begin as opacity-limited fragments in 
dynamically-unstable multiple systems and are ejected before 
they can accrete enough gas to become stars.  In no case does
a dense core collapse to form a single brown dwarf or a 
binary brown dwarf system alone.  This general mechanism has been
proposed by Reipurth \& Clarke \shortcite{ReiCla2001} and
Watkins et al.\ \shortcite{Watkinsetal1998b}, but the main
processes involved in the formation of the unstable multiple systems 
and the resulting properties of the brown dwarfs could only be 
conjectured.  

By tracing the evolution of the brown dwarfs in our calculation back 
to their fragmentation out of the gas, we find that
brown dwarfs form in two different ways.  Roughly three
quarters (14 of the 18 definite brown dwarfs) form via the 
fragmentation of gravitationally-unstable circumstellar discs 
(e.g.\ Figure 1).  The discs fragment due to rapid accretion 
(e.g.\ Bonnell 1994; Bonnell \& Bate 1994a,b; Whitworth et al.\ 1995; 
Burkert, Bate \& Bodenheimer 1997) and/or tidal-perturbations during
stellar encounters (Boffin et al.\ 1998; Watkins et al.\ 1998a,b).
The fragments destined to become brown dwarfs then interact
with the stars and brown dwarfs to which they are bound 
until they are ejected from the dense gas, limiting their masses 
to be substellar.  The remaining brown dwarfs (4 of the 18) originate 
from the fragmentation of dynamically-collapsing, filamentary 
molecular gas (e.g.\ Figure 2; Bonnell et al.\ 1991).  
The fragments initially form on 
their own, but quickly fall into unstable multiple systems and 
are ejected from the dense gas before they can attain stellar masses.
This latter mechanism is similar to that discussed 
by Reipurth \& Clarke \shortcite{ReiCla2001}.

Although the initial thermal Jeans mass in the cloud
is 1 M$_\odot$, the gas in the circumstellar discs and 
dynamically-collapsing gas is at a much higher density.  
Thus, the local Jeans mass and Jeans length are much smaller 
(both scale with density
as $\rho^{-1/2}$).  Consequently, when gravitational
collapse occurs in such an environment, the opacity-limited 
fragment that forms has a very limited reservoir of gas from
which to accrete.  In order to attain a large mass it must
accrete most of its mass from distances much greater than the
Jeans length of the gravitationally unstable clump from which
it formed.  However, since it is in a multiple system it must
compete for the gas (e.g.\ Bonnell et al.\ 1997).  
Furthermore, in an unstable multiple
system, the most frequently ejected object is that having the
lowest mass \cite{vanAlbada1968}.  These two constraints frequently 
conspire to eject a low-mass fragment from the dense gas 
in which it forms before it has accreted enough gas to become a star.

If brown dwarfs are produced by ejection from multiple systems, 
one might expect the fraction of binary brown dwarfs to be 
low since the binary must be ejected without being disrupted.
However, a close binary brown dwarf might be 
ejected from a multiple system and avoid disruption if
the other components of the system were widely separated.

This calculation does not produce any definite binary brown dwarfs,
but at the end of the calculation there is a close binary brown 
dwarf within an unstable multiple system.  The system 
consists of 7 objects.  The close binary brown dwarf
(semimajor axis 6 AU) is in a hierarchical binary 
(semimajor axis 84 AU, eccentricity 0.20) with a close stellar 
binary (semimajor axis 7 AU).  This quadruple system orbits 
a triple system at 330 AU with an eccentricity of 0.18.  
The triple system consists of a close (9 AU) stellar binary orbited 
by a brown dwarf companion (semimajor axis 90 AU, eccentricity 0.82).  
This system is dynamically unstable and 
will undergo further evolution.  Unfortunately, we are unable to 
follow it any further.  Since the binary brown 
dwarf is very close, it is possible that it will survive the dissolution
of this multiple system.  Even so, this would 
result in only one binary brown dwarf system and $\approx 20$ single 
brown dwarfs.  We therefore conclude that the formation of close 
binary brown dwarfs may be possible, but the fraction of 
brown dwarfs with a brown dwarf companion is very low ($\lsim 5$\%).

Observationally, the frequency of brown dwarf binaries is not yet clear.
Reid et al.\ (2001) find that 4 out of 20 brown dwarf primaries have 
companions giving a fraction of $\approx 20$\%.  However, this survey 
is magnitude limited rather than volume limited and is therefore likely to
overestimate the true frequency of brown dwarf binaries.  It is 
interesting to note
that none of the binary brown dwarf systems currently known have projected 
separations $> 10$ AU (Reid et al.\ 2001).  This is consistent with
their having survived ejection from unstable multiple systems.

Similarly, the survival of large discs around brown dwarfs 
is not an obvious outcome of the ejection mechanism of brown
dwarf formation.
Of the 18 definite brown dwarfs, most have
close dynamical encounters ($\lsim 20$ AU for 14 of the 18).
Of the remaining four, three are ejected before they can accrete 
gas with the high specific angular momentum required to form large discs.
Only one brown dwarf is ejected with a resolved disc 
(radius $\approx 60$ AU).  The other brown dwarfs are likely 
to possess smaller discs, but in this calculation we are unable 
to resolve discs with radii $\lsim 10$ AU (Section 2.2).  
Muench et al.\ \shortcite{Muenchetal2001} find $\approx 65$\% of 
brown dwarfs in the Orion Trapezium cluster have infrared excesses
indicative of discs.
Our results show that if brown dwarfs are formed by the ejection 
mechanism, most of these discs should have radii $\lsim 10$ AU.

Most brown dwarfs in our calculation are produced via gravitational 
instabilities in massive discs.  The ease with which disc 
fragmentation occurs depends primarily on the rate at which 
it accretes mass from the surrounding cloud (Bonnell 1994) 
and the disc's equation of state (e.g.\ Pickett et al.\ 2000).  
The density of the gas in the discs that fragment to form brown
dwarfs is high enough that the gas is in the $\eta=7/5$ regime
(Section 2.1).  Thus, the gas resists fragmentation far more than 
it would if an isothermal equation of state was used, although 
our equation of state does not include heating from shocks.  
On the other hand, because the flattened disc geometry may allow 
more rapid cooling than a spherically-symmetric geometry, real 
discs may be cooler and more unstable than those 
we model here.  Hence, the number of brown dwarfs may be even 
greater.  Another factor is that stars and brown dwarfs cannot 
merge in our calculation.  Mergers may reduce the final number of
brown dwarfs.  In summary, a more definitive 
prediction will have to wait until a large-scale calculation 
is performed with radiative transfer and even higher resolution.
For the present, we have demonstrated that the 
fragmentation of a turbulent molecular cloud is capable of 
forming similar numbers of brown dwarfs and stars.  
This is in agreement with observational surveys of brown 
dwarfs in the solar neighbourhood (Reid et al.\ 1999)
and in clusters (e.g.\ Luhman et al.\ 2000 and references therein).

\section{Conclusions}

We have presented results from the first calculation to 
demonstrate the formation of brown dwarfs from the 
fragmentation of a turbulent molecular cloud.  The 
calculation resolves the opacity limit for fragmentation
and follows numerous brown dwarfs until accretion onto them
has ceased.

We find that the star-formation process produces 
roughly equal numbers of stars and brown dwarfs, in agreement 
with recent observations.  Examining the mechanisms by which the 
brown dwarfs form, we find that they are the result of the 
ejection of fragments from the dense gas in which they form by 
dynamical interactions in unstable multiple systems.  This occurs 
before they are able to accrete to stellar masses.
Three quarters of the brown dwarfs fragment out of 
gravitationally-unstable discs, while the remainder form in 
collapsing filamentary flows of high-density gas.  

The calculation indicates that close binary brown dwarf systems 
(separations $\lsim 10$ AU) might be able to survive the ejection 
process.  However, such systems should be very rare 
(frequency $\lsim 5$\%) because of the close dynamical interactions 
that are involved in the ejection of brown dwarfs.  Similarly,
large discs (radii $\gsim 10$ AU) around young brown dwarfs 
should also be rare (frequency $\approx 5$\%).  Observations 
show that close binary brown dwarfs and circumstellar discs 
surrounding brown dwarfs do exist, but more detailed surveys are 
required to test the predictions made here.

\section*{Acknowledgments}

The computations reported here were performed using the UK Astrophysical 
Fluids Facility (UKAFF).  We thank the referee for prompting 
several improvements to the letter.

\end{document}